\documentclass[twocolumn,prl,showpacs]{revtex4}

\usepackage{graphicx}
\usepackage{rotating}
\usepackage{amsmath}
\usepackage{amsfonts}
\usepackage{amssymb}
\usepackage{enumerate}
\usepackage{longtable}
\usepackage{dcolumn}
\usepackage{bm}

\begin{document}

\newcommand{\be}{\begin{equation}}
\newcommand{\ee}{\end{equation}}
\newcommand{\bn}{\begin{eqnarray}}
\newcommand{\en}{\end{eqnarray}}

\title{On Field-Induced Quantum Criticality in $YbRh_{2}Si_{2}$.   
}

\author{M. S. Laad}

\affiliation{Max-Planck Institut f\"ur Physik Komplexer Systeme, 
N\"othnitzer Strasse 38, D01187 Dresden, Germany}

\date{\today}

\begin{abstract}
 The quantum critical point (QCP) in $YbRh_{2}Si_{2}$ is an enigma for the
itinerant view of QCP. In an alternative view, 
this QCP is intimately linked to the selective Mott localization of the heavy $f$ electrons.  Following a perusal of this unusual QCP,
   I study an Extended Periodic Anderson Model (EPAM)
within DMFT.
A quantum phase transition (FQPT),
accompanied by a rapid change 
in the Fermi volume, is found near the quantum-critical end-point of the 
selective Mott transition in the $f$-electron sector.  The theory accounts for
a wide range of unusual, singular non-Fermi liquid features exhibited at this QCP in 
$YbRh_{2}Si_{2}$ in a natural way. 
\end{abstract}


\maketitle


Quantum Criticality in solids and the potential to ``tune in'' to new,
novel phases of matter in their proximity underpins a large component 
of modern condensed matter physics research~\cite{[gil],[gegenwart]}.  
The rare-earth compound $YbRh_{2}Si_{2}$ is a particularly enigmatic case in 
point, exhibiting a ``strange'' non-FL ``phase'' fanning out from a 
$T=0$ QCP separating antiferromagnetic (AF) ordered and heavy Fermi liquid 
(HFL) phases.  $YbRh_{2}Si_{2}$ is tuned to the QCP by minute magnetic field
or chemical substitution.  Experimental data indicate that the Hertz-Moriya-Millis (HMM) scenario~\cite{[millis]} does not account for the unusual 
responses in $YbRh_{2}Si_{2}$.  An alternative view posits that these are 
associated with the destruction of the Kondo effect itself, either by inter-site
 RKKY interactions~\cite{[gegenwart],[senti]}, or by selective localization of 
$f$ electrons~\cite{[pepin]}.   
Given that the ``standard model'' of $f$-band systems, the Periodic 
Anderson Model (PAM), emphasizes the HFL aspect driven by {\it quasilocal} Kondo screening~\cite{[3]}, these new observations call for  
mechanisms which destabilize Kondo singlet formation.  In spite of vigorous 
attempts~\cite{[gegenwart],[pepin]}, the problem is far from
``being solved''.    

I begin by recapitulating salient features of the FQCP in $YbRh_{2}Si_{2}$:

(i) $dc$ resistivity, $\rho(T)\simeq AT$ over three decades in $T$~\cite{[steg1]},
(ii) specific heat, $C_{V}(T) \simeq T^{0.6}$ at very low 
$T$~\cite{[gegenwart]},
(iii) anomalously slow (in frequency, $\omega$) decay of optical conductivity, with linear-in$\omega$ scattering rate, $\tau^{-1}(\omega)\simeq \omega^{\alpha}$ with $\alpha\simeq 1$~\cite{[steg3]},
(iv) strongly $T$-dependent Hall constant, $R_{H}(T)$, and cot$\theta_{H}(T)\simeq C_{1}T^{2}+C_{2}$, and
(v) rapid change in the low-$T$ value of $R_{H}$ across the FQCP, extrapolating to a jump as $T\rightarrow 0$~\cite{[steg5]}, suggesting a rapid change in the Fermi surface (FS) across the FQCP,
(vi) static magnetic susceptibility, $\chi({\bf q}=0,T,B=0) \simeq T^{-0.6}$ for
$T>0.3$~K, along with large Korringa ratio~\cite{[steg7]}, 
indicating very strong {\it ferromagnetic} correlations close to the
FQCP.  And $\chi(B) \simeq (B-B_{c})^{-0.6}$ scales with the $A$-co-efficient of
the $T^{2}$ term in $\rho(T)$ in the HFL regime, 
(vii) NMR derived Knight shift, $K_{s}(T,B)$ and the relaxation rate, $1/T_{1}T$ scale with $\chi({\bf q}=0,T,B=0)$.

  (i),(iii) and (iv) are reminiscent of what is seen in high-$T_{c}$ cuprates 
in their ``normal'' state.  Recently, (v) has also been seen in cuprates near
optimal doping~\cite{[louis]} All these behaviors are at odds with 
the HMM theory~\cite {[millis]}, which predicts markedly different behavior~\cite{[si]}.  So the elucidation of (i)-(vii) in a single theoretical picture 
remains a challenge.

 Here, I address these issues by proposing a modified  
 PAM with extended $f$-hopping and hybridisation,
 as well as a direct coulomb interaction between 
the $f$ electrons and conduction ($c$) electrons, dubbed Extended-PAM (EPAM).
I study this EPAM using DMFT,
showing how the non-FL behavior along a curve in parameter space is understood
as a selective Mott localization, and discuss how (i)-(vii) naturally follow
therefrom.
To the extent that this non-FL 
behavior is tied to $f$-Mott physics,
single-site DMFT should capture the relevant physics.  I will also show how 
this non-FL state is unstable to either AF, or to a heavy FL (HFL) away from 
this curve, at $T=0$.
  
 The Hamiltonian is $H=H_{0}+H_{1}$, with the band part described by

\be
H_{0}=-t_{f}\sum_{<i,j>,\sigma}f_{i\sigma}^{\dag}f_{j\sigma} -
t_{p}\sum_{<i,j>,\sigma}c_{i\sigma}^{\dag}c_{j\sigma} + V_{fc}\sum_{<i,j>,\sigma}f_{i\sigma}^{\dag}c_{j\sigma} 
\ee

and the local part, by

\be
H_{1}= U_{ff}\sum_{i}n_{if\uparrow}n_{if\downarrow} + U_{fc}\sum_{i,\sigma,\sigma'}n_{if\sigma}n_{ic\sigma'}  
+ \epsilon_{f}\sum_{i}n_{fi}
\ee 
I take the $c$-band centered around $E=0$ and consider
$U_{ff}=\infty$ (valid for $f$ shells), so the $f$ electrons are 
projected fermions, $X_{if\sigma}=(1-n_{if-\sigma})f_{i\sigma}$,
satisfying $[X_{i\sigma},X_{j\sigma'}^{\dag}]_{+}=\delta_{ij}\delta_{\sigma\sigma'}(1-n_{if,-\sigma})$. 
Using the Gutzwiller approximation, $X_{if\sigma}=q_{\sigma}f_{i\sigma}$ with $q_{\sigma}=(1-n_{f})/(1-n_{f\sigma})$ and $n_{f\sigma}=(1/N)\sum_{i}\langle f_{i\sigma}^{\dag}f_{i\sigma}\rangle$.
This implies $(t_{f},\epsilon_{f})\rightarrow q_{\sigma}^{2}(t_{f},\epsilon_{f})$ and 
$V_{fc}\rightarrow q_{\sigma}V_{fc}$ in what follows, and corresponds to the
slave-boson mean-field theory (SB-MFT), yielding a narrow, coherent $f$ band
with a width $W\simeq k_{B}T_{K}^{mf}$~\cite{[pepin]}, the mean-field Kondo 
scale.  Consistent with LDA calculations~\cite{[gertrud]}, I take 
$\epsilon_{f}\simeq E_{F}$.  Below, I investigate the fate of this SB-MFT 
Kondo scale in presence of strong, quantum fluctuations of the $f$ occupation, 
caused by the competition between mean-field coherence, 
$T_{K}^{mf}(V_{fc},(t_{f,p}/U_{ff}))$ and incoherence, driven by $U_{fc}$.
  
  I start by splitting the $V_{fc}$ term as
$(V_{fc}-\sqrt{t_{f}t_{p}})\sum_{<i,j>,\sigma}(f_{i\sigma}^{\dag}c_{j\sigma}+h.c)
+\sqrt{t_{f}t_{p}}\sum_{<i,j>,\sigma}(f_{i\sigma}^{\dag}c_{j\sigma}+h.c)$
and consider $H=H_{0}+H_{1}$ with $V_{fc}^{(1)}=\sqrt{t_{f}t_{p}}$ to begin 
 with.
Using $a_{i\sigma}=(uf_{i\sigma}+vc_{i\sigma}), b_{i\sigma}=(vf_{i\sigma}-uc_{i\sigma})$
with $u=\sqrt{t_{f}/(t_{f}+t_{p})}, v=\sqrt{t_{p}/(t_{f}+t_{p})}$.
it is easy to see that $H=H_{0}+H_{1}$ is
$H_{0}=-t\sum_{<i,j>,\sigma}(a_{i\sigma}^{\dag}a_{j\sigma}+h.c)$ and
$H_{1}= U_{fc}\sum_{i,\sigma,\sigma'}n_{ia\sigma}n_{ib\sigma'} + 
\epsilon_{f}\sum_{i,\sigma}[n_{ia\sigma}+n_{ib\sigma}+(a_{i\sigma}^{\dag}b_{i\sigma}+h.c)]$.

  This is the spin $S=1/2$ Falicov-Kimball model (FKM) with a 
{\it local} hybridisation term, which is finite whenever $\epsilon_{f}\ne 0$.   
Remarkably, when $\epsilon_{f}=0$, this 
reduces to the pure $S=1/2$ FKM!  Below, we show how $(\epsilon_{f}=0,V_{fc}=\sqrt{t_{f}t_{p}})$ 
separates {\it two}, different metallic phases.  

  First, at $\epsilon_{f}=0$, we see that $[n_{ib},H]=0$ for {\it each} $i$,
implying a {\it local} $U(1)$ invariance of $H$: local configurations with
$n_{b}=0,1$ are rigorously degenerate.  This is {\it exactly} the condition for
having singular $b$-``number'' fluctuations.  As is known~\cite{[si1]}, the symmetry
unbroken metallic phase is consequently {\it not} a FL, but is dominated by a 
superposition of one-particle ($n_{b}=0$) and two-particle ($n_{b}=1$) states
at low energy.  The $a$-fermion propagator 
is, assuming a lorentzian unperturbed DOS with half-width $W$ for analytical clarity, very simple, showing ``upper'' and ``lower'' Hubbard bands, with a 
pseudogap at $E_{F}(=0)$:

\be
\rho_{a}(\omega)=\frac{1-n_{b}}{\omega^{2}+W^{2}} + \frac{n_{b}}{(\omega-U_{fc})^{2}+W^{2}}
\ee
The corresponding $b$-fermion propagator has branch cut
singular behavior precisely at $E_{F}(=0)$, leading to singularities in the 
local,  one- and two-particle responses: $\rho_{b}(\omega) \simeq\theta(\omega)|\omega|^{-(1-\alpha_{0})}$ 
and $\chi_{ab}"(\omega)=\int dt e^{i\omega t}\langle
a_{i\sigma}^{\dag}b_{i\sigma}(t);b_{i\sigma}^{\dag}a_{i\sigma}(0)\rangle\simeq \theta(\omega)|\omega|^{-(2\alpha_{0}-\alpha_{0}^{2})}$.  
Here, $\alpha=(1/\pi)$tan$^{-1}(U_{fc}/W)$ is the so-called $s$-wave phase
shift of the Anderson-Nozieres-de Dominicis (AND) X-Ray Edge (XRE) problem.  Obviously, the FL quasiparticle 
weight, $Z=0$, and the $\omega, T$ dependence of physical quantities will
be governed by power-law responses.  Notice that, in the $(f,c)$ basis,
divergence of $\chi_{ab}"(\omega)$ corresponds to extended (cf. non-local hybridisation),
singular quantum fluctuations of the $f$
occupation: it is precisely these fluctuations which
destroy FLT at $\epsilon_{f}=0$ via the 
AND orthogonality catastrophe (OC)~\cite{[12]}.  Given that the $4f_{7/2}$ level
 hybridizes with {\it two} ``$c$'' bands in reality~\cite{[gertrud]},  we get
the OC exponent, $\alpha=2\alpha_{0}$.

  In our FKM with $\epsilon_{f}=0$, and at $T=0$, 
the total fermion number, $n=n_{a}+n_{b}=n_{f}+n_{c}$, jumps discontinuously 
from $n_{-}=(1/2)+(1/\pi)tan^{-1}(U_{fc}/2W)$ to $n_{+}=(3/2)-(1/\pi)tan^{-1}(U_{fc}/2W)$ for a range of densities, $n$, near unity.  This corresponds to a 
sudden jump in the $b$-occupation, $n_{b}$, 
giving a {\it first} order 
``valence'' transition as $\epsilon_{f}$ is tuned through $\mu(=0)$. 
At finite $T$, this line of first-order transitions ends at a 
second order critical end-point (CEP), and
$n_{b}$ varies very rapidly over an energy scale $O(k_{B}T)$ 
around $\epsilon_{f}=0$, 
extrapolating to a jump $T=0$.  Notice that this jump in $n_{b}$ depends on 
$U_{fc}/t$, which we choose henceforth to be such that this jump is vanishingly
small~\cite{[miyake]} at $T=0$, giving a quantum critical end-point (QCEP). 
This explicitly shows the link between emergence of singular non-FL behavior 
and the selective Mott localization of the $b$-electrons. 
{\it We emphasize that, with finite $t_{f},V_{fc}$,  both, the $f$- and
$c$-fermions remain mobile: only their combination, $b_{\sigma}=(vf_{\sigma}-uc_{\sigma})$, is localized}.  Recent slave boson approaches 
have proposed this ``selective Mott transition'' of $f$-electrons in the context of the QCPs in RE systems~\cite{[pepin]}.  Our work is a concrete, DMFT-based,
realization of the ``selective Mott'' QPT, with a {\it non-local} hybridization.
In contrast to earlier work~\cite{[pepin]}, nFL behavior here arises from the
AND-OC in the corresponding impurity problem as $V_{fc}$ is varied across a 
critical value, $V_{fc}^{(1)}=\sqrt{t_{f}t_{p}}$. 

  We now show how the unique observations at the FQCP in $YbRh_{2}Si_{2}$ are
understood as a consequence of the AND-OC derived above.
  The singularity in the $b$-DOS implies that their contribution to thermodynamic responses dominates that of the ``itinerant'' $a$-fermions.
Hence, the low-$T$ specific heat is 

\be
C_{el}(T) \simeq T.lim_{\eta\rightarrow 0}Im[G_{bb}(\omega+i\eta)]|_{\omega=T}\simeq T^{\alpha}.  
\ee
giving the $\gamma$ co-efficient as $\gamma(T)\simeq T^{-(1-\alpha)}$; actually, goes like $T^{-(1-\alpha)}$log$(T/T_{coh})$ as the QCP is approached.  
The log- factor comes from seeing that the DOS is approximately a lorentzian 
with a maximum varying like $T^{-(1-\alpha)}$ and a full width at half-maximum
equal to $\alpha\pi T$.  When $E_{F}(=0)$ lies within this peak, we can write
$E_{F}(T)=-\alpha\pi T$, whence the asymptotic form of $C_{el}\simeq T^{-(1-\alpha)}$log$(T/T_{coh})$ follows.
The 
entropy is then directly obtained as $S(T)=\int_{0}^{T}\gamma(T')dT'\simeq T^{\alpha}$.  With $\alpha_{0}=0.3$, we thus find that both $C(T),S(T)$ vary as $T^{\alpha}$ with $\alpha=0.6$, in nice agreement with observations in 
the nFL regime as a function of $T$~\cite{[gegenwart]}. 

  What about transport?.  In the impurity limit (note that the lorentzian unperturbed DOS in DMFT will not modify the ``impurity'' result), we have $G_{a0}(\tau)=(\pi T\rho_{0}/$sin$(\pi T\tau))$ and $G_{b0}(\tau)=sgn(\tau)/2$, whence the respective self-energies are $\Sigma_{a}(\tau)=U_{fc}^{2}G_{a0}(\tau)G_{a0}(\tau)G_{a0}(\tau)$ and
$\Sigma_{b}(\tau)=U_{fc}^{2}G_{a0}(\tau)G_{a0}(\tau)G_{b0}(\tau)$.  
Direct evaluation followed by Fourier transformation then gives
$\Sigma_{b}(i\omega_{n})=-i(U_{fc}\rho_{0})^{2}[\omega_{n}($ln$(E_{F}/T)-\Psi(\omega_{n}/2\pi T)- \pi T]$ and $\Sigma_{a}(i\omega_{n}) \simeq (\omega^{2}+\pi^{2}T^{2})$.  The $dc$ resistivity within DMFT is then $\rho_{dc}(T) \simeq (m/ne^{2})$Im$\Sigma_{b}(\omega)|_{\omega=T}\simeq AT$, i.e, it is linear in $T$.  The
optical scattering rate will also be linear in $\omega$.  This is exactly in 
accord with observations near the QCP in 
$YbRh_{2}Si_{2}$~\cite{[steg1],[steg3]}.  Interestingly, with log-singularities
in $\Sigma_{b}$ above (which already imply $Z=0$), higher-order terms, which 
must be carefully examined, lead precisely to the {\it branch cut} singular 
structure~\cite{[yulu]} for $\rho_{b}(\omega)$, characteristic of the ``lattice X-ray edge'' problem found in DMFT.  
  
To proceed, observe that the impurity model corresponding to $H_{FKM}$ can be
bosonized in each radial direction centered around the ``impurity'' site~\cite{[13]}.  
For general band-filling, $n=n_{a}+n_{b}\ne 1$ per site, the umklapp terms 
from $U_{fc}$ are irrelevant and hence ignored.  The bosonized Lagrangian then
describes a collection of {\it non-interacting} charge- and spin density 
collective modes:
\be
L_{0}'=\sum_{\rho,\sigma}\frac{u_{\rho,\sigma}}{2}\int [K_{\rho,\sigma}\Pi_{\rho,\sigma}^{2}(r)+\frac{1}{K_{\rho,\sigma}}(\partial_{r}\phi_{\rho,\sigma}(r))^{2}]dr
\ee
and
$L_{0}"=\frac{g}{\pi u_{\rho}}\sum_{\rho}\int \partial_{r}\phi_{\rho}(r)dr$.  Here,
$g,u_{\rho,\sigma},K_{\rho,\sigma}$ are 
explicit functions of $U_{fc}/t$.  Thus, interactions simply ``shift'' 
the {\it charge} bosonic 
modes relative to their free values.    
 Introducing the usual symmetric-antisymmetric (charge-spin) combinations of
$\phi_{\rho,\sigma}(r)$, we see that the antisymmetric (spin) channel
 completely decouples from the charge channel: 
a kind of high-dimensional 
{\bf spin-charge separation}!  This has been strongly emphasized by Anderson~\cite{[12]} in the cuprate context, and has important consequences, detailed 
below. 

 $\epsilon_{f}\ne 0$ has two effects: (i) it moves the 
$b$-fermion level away from $E_{F}$, and, 
(ii) finite $a$-$b$ hybridisation 
generates a finite, but heavy $b$-fermion mass, due to recoil in the XRE 
problem~\cite{[14]}, giving a small ``coherence scale'', $\epsilon_{rec}=k_{B}T_{coh}$,
 below which HFL behavior obtains in the 
lattice model.  The FL quasiparticle overlap, $Z \simeq e^{-C(t=\infty)}$,  
with $C(t)=2U_{fc}^{2}\int \frac{\chi_{ab}"(\omega)}{\omega^{2}}(1-cos(\omega t))d\omega$.  This gives $Z \simeq exp[U_{fc}^{2}(ln(\kappa)/(1-\kappa^{2}))]$.  Here, $\kappa=m_{a}/m_{b}$, with $m_{a}$ the band mass of
the $a$-fermion and $m_{b}$ the heavy mass of the $b$ fermion. 
Hence $T_{coh} \propto Z$ (note the difference from the SBMFT 
scale, $T_{K}^{mf}$)
increases with $\epsilon_{f}$, as indeed observed in the
region to the right of the FQCP.  In $D=\infty$, the
relevant hybridization, $\epsilon_{f}\sum_{i,\sigma}(a_{i\sigma}^{\dag}b_{i\sigma}+h.c)$ implies that the one-electron
DOS will show a narrow, low-energy FL resonance, with upper/lower Hubbard 
bands at high energies, as is known~\cite{[si1]}.
  Away from $V_{fc}=\sqrt{t_{f}t_{p}}$, the
term $\sum_{<i,j>,\sigma}\delta V_{fc}(f_{i\sigma}^{\dag}c_{j\sigma}+h.c)$ also causes one-particle {\it intersite} hybridisation between the 
$a,b$ fermions.
This again gives the $b$-fermions a finite mass and  
results in another HFL with $T_{coh} \propto Z<<1$.  Low-energy responses are
then those of a HFL with $Z<<1$: an enhanced $\gamma=C_{el}(T)/T, \chi(T)=\chi_{0}, \rho_{dc}(T)=\rho_{0}+AT^{2}$, etc, followed by a {\it smooth} crossover to
the non-FL response found for $V_{fc}^{(1)}=\sqrt{t_{f}t_{p}},\epsilon_{f}=0$.  

Using the bosonized form, Eq.(7), above allows further progress in the nFL 
regime.
  Expressing the transverse spin correlation function as an average over the 
phase variables permits its 
evaluation using $L_{0,\sigma}$.  The result, following~\cite{[gogolin]}, is

\be
\chi^{+-}({\bf q},\omega)=\frac{A}{T^{K_{\sigma}^{-1}-K_{\rho}}}F(\frac{\omega}{T})
\ee
with $K_{\rho}=\sqrt{v_{F}/(v_{F}+U_{fc})}$ and $v_{F}=2t$ the Fermi velocity.
$F(x)$ is a scaling function, $\simeq x$ for $x<<1$ and $\simeq 1$ for $x>>1$.
And $K_{\sigma}=1$ for the SU$(2)$ invariant case, but $K_{\sigma}<1$ including spin-orbit ($s-o$) coupling effects. 
This immediately yields the power-law $T$-dependence of the NMR relaxation 
rate as

\be
\frac{1}{T_{1}}=\frac{T}{\omega}\sum_{{\bf q}}Im \chi^{+-}({\bf q},\omega) \simeq T^{-(K_{\sigma}^{-1}-K_{\rho})}
\ee
The uniform spin susceptibility follows as $\chi({\bf q}=0,T)\simeq T^{-(K_{\sigma}^{-1}-K_{\rho})}$.  In DMFT, the singular-in-$\omega$ part of $\chi({\bf q},\omega)$ is independent of ${\bf q}$.  This explains why {\it both} $T_{1}^{-1}(T)$ and Knight shift, $K_{s}(T)$, scale like $T^{-(K_{\sigma}^{-1}-K_{\rho})}$ like $\chi({\bf q}=0,T)$.
  With a reasonable choice of $U_{fc}/t$, we get $K_{\rho}=0.4$, leading to very good agreement with the host of power-law behaviors found in the magnetic response near the FQCP in $YbRh_{2}Si_{2}$.  In particular, with the choice $K_{\sigma}=1$,   
$\chi({\bf q}=0,T), T_{1}^{-1}(T), K_{s}(T)$ all follow a $T^{-0.6}$ law.  Further, taking $\alpha_{0}=0.3$ (see above), we find 
$\chi/\gamma \simeq T^{-0.2}$.  {\it Assuming} $b/T$ scaling, where
$b=(B-B_{c})$ is the distance from the critical field, this implies
$\chi(b)/\gamma(b) \simeq b^{-0.2}$, and that $\chi(b)\simeq b^{-0.6}, \gamma(b)\simeq b^{-0.4}$ near the FQCP.  Further, with $A(b)\simeq 1/b$~\cite{[gegenwart]}, we find
that the Woods-Saxon ratio, $A/\gamma^{2}\simeq b^{-0.2}$ and $A/\chi^{2}\simeq b^{0.2}$, which is weakly $b$-dependent and saturates at ``higher'' $b$.  
{\it All} these are experimentally seen~\cite{[steg7]}.  While $\alpha_{0}=0.3$ as found in our {\it model} DMFT may change somewhat in a truly ``first principles'' theory, the qualitative theory-experiment agreement is compelling.  

  The higher-$D$ spin-charge separation implied by the bosonized form of the 
impurity model also leads to consistency with the magnetotransport results:
the Hall relaxation rate is now controlled by spinon-spinon scattering, leading 
to cot$\theta_{H}(T)\simeq c_{1}T^{2}+c_{2}$~\cite{[12]}.  
With $\rho_{dc}(T)=A_{c}T$, 
we get the Hall resistivity, $\rho_{xy}(T)\simeq T^{-1}$.  {\it Both} are in good agreement with experiment.

  We now turn to the evolution of the FS across the FQCP in $YbRh_{2}Si_{2}$.  Hall data suggest an abrupt reconstruction of the FS across the FQCP as $T\rightarrow 0$.  Thus, a large FS in the HFL regime, also seen in dHvA work~\cite{[julian]}, abruptly goes over to a small FS on the AF side.  
Within DMFT, in the symmetry-unbroken metallic phase(s), the shape {\it and} size
of the FS is {\it not} affected by interactions, since the self-energy is purely local: $\Sigma_{a,b}(k,\omega)=\Sigma_{a,b}(\omega)$.  Exactly at the FQCP,
i.e, at $V_{fc}=\sqrt{t_{f}t_{p}},\epsilon_{f}=0$, the 
FS will be a single sheet with a volume corresponding
to the $a$-fermion number.  To the right of the FQCP, the finite $\delta V_{fc},\epsilon_{f}$ gives a finite $a-b$ hybridization, giving a HFL metal, as found
above.  In this regime, DMFT studies on the PAM with a relevant hybridization
yield a large FS containing {\it both}, the lighter $a$- as well as the heavy
$b$-fermions.  Thus, the abrupt change in the FS volume is intimately
 linked with the selective Mott localization of the $b_{\sigma}=(vf_{\sigma}-uc_{\sigma})$ fermions; they ``decouple'' from the FS at the FQCP.  
This occurs exactly at the point where the FL coherence 
scale vanishes, giving a non-FL metal with low-energy singular responses.  

  What about AF order?  For small $\delta V_{fc},\epsilon_{f}<<1$, {\it two-particle}
 processes, generated to second order in $\delta V_{fc},\epsilon_{f}$, are more
relevant than the one-particle $a-b$ hybridization.  These processes couple two
``impurities'', and lead to two-particle instabilities,  
as in 
coupled $D=1$ Luttinger liquids~\cite{[gogolin]}.  
To this order, extra terms,   
$H_{res}^{(2)}\simeq -\lambda^{2}\sum_{<i,j>}a_{i\sigma}^{\dag}b_{i\sigma}b_{j\sigma'}^{\dag}a_{j\sigma'}$, with $\lambda^{2}\simeq O((\delta V_{fc})^{2}/U_{fc})$, are generated in $H$.  
In $D=\infty$,
these are decoupled as 
$H_{res}^{eff}=-\lambda^{2}\sum_{<i,j>,\sigma}(M_{ab}a_{i\sigma}^{\dag}b_{i\sigma}+ M_{b}a_{i\sigma}^{\dag}a_{j,-\sigma}+h.c)$.  Solving $H=H_{FKM}+H_{res}^{(2)}$
within DMFT should yield an AF metallic phase~\cite{[amg]}: it will have the 
same symmetry as the AF phase in an ``iitinerant'' view, since {\it both} have $M_{a}=\langle a_{i\sigma}^{\dag}a_{j-\sigma}\rangle >0$.  However, I choose a different 
route.  Bosonizing these terms in $H_{res}^{eff}$, 
 the second term, corresponding to AF order, generates a cosine term 
in the bosonized Lagrangian for the spin sector: $L_{\sigma}^{int}=g_{1}$cos$(\beta\phi_{\sigma})$, with $\beta=\sqrt{8\pi K_{\sigma}}$.   
With spin-orbit interaction, $K_{\sigma}<1$, and the
Lagrangian in the spin sector,

\be
L_{\sigma}=L_{0,\sigma}+g_{1} \int cos(\beta\phi_{\sigma})dr
\ee

is a quantum sine-Gordon model with a relevant cosine term.  This 
leads to an AF ordered state, corresponding to a finite
expectation value of the $\phi_{\sigma}$ field: $\langle\phi_{\sigma}\rangle >0$~\cite{[gogolin]}.  
Thus, AF order here results as a particle-hole instability of the 
singular, non-FL metal derived above in the spin channel, rather than from 
a band FS instability, as would be the case in ``conventional'' cases where FS
nesting features in a FL metal give itinerant magnetism: the latter picture cannot account for power-law responses seen at the QCP in $YbRh_{2}Si_{2}$.  
Thus, in our EPAM, at the FQCP, selective $b$-fermion localization 
 permits AF to arise simply due to ``inter-impurity'' (corresponding to onset 
of RKKY-like) $b-b$ local moment correlations induced via ``itinerant''
$a$-fermions.  Obviously, the FS now has a small volume, containing only the
``itinerant'' $a$-fermions.

  At $V_{fc}^{(1)}$, the AND-OC will {\it always} occur in the symmetry-unbroken
 metallic phase in {\it any} $D<\infty$.  Thus, we expect that our findings 
will survive inclusion of non-local correlations beyond DMFT.  
Also, the $f$-electrons are {\it never} strictly localized: only the
 $b$-combination localizes at $V_{fc}^{(1)}$. 
  An ab-inito theory for $\alpha=0.6,K_{\rho}=0.4$ and $K_{\sigma}$ 
used here is hard: here, we have employed plausible $U_{fc}/t=10$ (this is 
the {\it only} free parameter in our model) values.  A truly
first-principles correlated program (e.g, LDA+DMFT) is required to {\it derive} them.  We plan to address this issue in future.

  In conclusion, a {\it local} QCP, triggered by the AND-OC~\cite{[pwa]}, 
is found in 
the DMFT solution of the EPAM as the model parameters are varied.  Using high-$D$ bosonization, non-FL responses with an uncanny resemblance to those found at
the FQCP in $YbRh_{2}Si_{2}$ are uncovered.  This QCP is unstable, either to a 
HFL, or to AF.  {\it All} these findings are in very good qualitative agreement with the $T-b$ phase diagram of $YbRh_{2}Si_{2}$, whose unconventional QCP is 
thence proposed to be of the local type, and associated with the selective Mott 
localization in the EPAM.  Our analysis is potentially applicable to other,
$d$- and $f$-electron based systems showing non-FL behaviors near the 
$T\rightarrow 0$ itinerant-localized transitions.

{\bf Acknowledgements}

  I thank Prof. G. Lonzarich for discussions and his suggestion to look closer at quantum critical end-point of the valence transition in $f$-band 
systems.

\end{document}